\documentclass{ws-procs9x6}

\usepackage{revsymb}

\def\dd{{\mathrm d}}
\def\ii{{\mathrm i}}
\def\ee{{\mathrm e}}

\begin{document}

\title{Noble gas, alkali and alkaline atoms interacting with a gold surface}

\author{Grzegorz \L{}ach 
and Maarten DeKieviet
}

\address{Physikalisches Institut der Universit\"{a}t Heidelberg, \\
Albert-Ueberle-Strasse 3-5, 69120 Heidelberg, Germany}

\author{Ulrich D. Jentschura
}

\address{Department of Physics, Missouri University of Science
and Technology,\\ Rolla, Missouri 65409-0640, USA}

\begin{abstract} 
The attractive branch of the interaction potentials with the surface of gold have
been computed for a large variety of atomic systems: the hydrogen atom, noble
gases (He, Ne, Ar, Kr, Xe), alkali atoms (Li, Na, K, Rb, Cs) and alkaline atoms
(Be, Mg, Ca, Sr, Ba). The results include highly accurate dynamic polarizabilities
for the helium atom calculated using a variational method and explicitly correlated
wavefunctions. For other atoms considered we used the data available in the literature.
The interaction potentials include both the effects of retardation of the electromagnetic
interactions and a realistic representation of the optical response function of gold
(beyond the approximation of a perfect conductor). An explicit comparison of our
result to the interaction between an atom and a perfect conductor is given.
\end{abstract}


\bodymatter

\section{Motivation}\label{aba:sec1}

In the last decade, there has been drastic experimental progress in the field
of (ultra-) cold collisions in atomic systems. These may involve interactions
among two atoms as well as those between an atom and a solid wall.
Among the latter, one of the experimental methods for studying the long
range ($z \gg a_0$), attractive part of the interaction potential makes use of
the phenomenon of quantum reflection. It was demonstrated, that scattering
cold atomic beams under grazing angles from well defined, single crystal
solid surfaces is especially useful for probing Casimir-Polder forces.
This is due to the fact that quantum reflectivity is particularly sensitive to the
shape of the potential, at distances where it is heavily modified by
retardation\cite{DrDK2003}. Other techniques involving the manipulation
of clouds of cold atomic gases, or condensates in the presence of a solid
wall\cite{HaCo2005} heavily depend on the  details of the interaction
potential. Despite the abundance of experimental results, the theoretical
analysis of the interaction potentials was often based on simple model potentials
with parameters chosen to fit the data.  Here, we provide accurate atom-surface
interaction potentials for the surface of gold based on ab-initio computed atomic dynamic
polarizabilities and a compilation of all available optical data in literature.

\section{Atom-surface interactions}
For distances where the exchange effects become negligible, the atom-surface
can be computed by considering a polarizable particle interacting with quantum
electromagnetic field fluctuations. In this approach, the solid  is treated as a
continuous medium having a frequency dependent permittivity $\epsilon(\omega)$.
The derivation of the dipolar term $V_1(z)$, dominant at the long-distance limit,
has been first performed by Lifshitz\cite{Li1955}, and the result when given in atomic
units ($m_e = e =\hbar =1$) reads:
\begin{equation}
\label{dipanydielectric}
V_1(z) =
-\frac{\alpha^3}{2\pi}\int_0^\infty\dd\omega\,\omega^3\,
\alpha_1(\ii \omega)
\int_1^\infty\dd\xi\,
\ee^{-2\alpha\xi\omega z}\,
{\rm H}(\xi,\epsilon(\ii\omega))\ ,
\end{equation}
where $\alpha$ is the fine structure constant and $\alpha_1(\ii \omega)$ is the 
dipole polarizability of the atom [for the fundamental
physical constants we take the CODATA\cite{MoTaNe2008} recommended values,
e.g., $\alpha=137.035999679(94)$]. The function ${\rm H}(\xi,\epsilon)$ is given by:
\begin{equation} \label{hfunction}
{\rm H}(\xi,\epsilon) =
(1-2\xi^2)\frac{\sqrt{\xi^2 + \epsilon-1} - \epsilon\,\xi}{\sqrt{\xi^2+\epsilon-1} +\epsilon\,\xi}
+\frac{\sqrt{\xi^2+\epsilon-1}-\xi}{\sqrt{\xi^2+\epsilon-1}+\xi} \,.
\end{equation}
This expression simplifies considerably in the limit of a perfect conductor
[$\epsilon(\omega) \to \infty$], for which case the potential becomes:
\begin{equation}\label{dipanycond}
V^{(\infty)}_1(z) =
-\frac{1}{4\pi}\int_0^\infty\dd\omega\,\alpha_1(\ii \omega)\,\ee^{-2\alpha\omega z}\,P^{(\infty)}_1(\alpha\omega z)\ ,
\end{equation}
where $P_1^{(\infty)}(z)=1+2z+2z^2$.
Another simplification takes place in the short-distance and long-distance limits.
For $z\to 0$, the interaction potential behaves as:
\begin{equation}
V_1(z) \mathop{\sim}^{z\to 0}-\frac{1}{4\pi z^3}\int_0^\infty\dd\omega\,
\alpha_1(\ii \omega)\frac{\epsilon(\ii\omega)-1}{\epsilon(\ii\omega)+1} 
\equiv  -\frac{C_3}{z^3} \,.
\end{equation}
The long-distance limit the interaction potential for a generic $\epsilon(\omega)$ is:
\begin{equation}
V_1(z)\mathop{\sim}^{z\to\infty}-
\frac{3\,\alpha_1(0)}{8\pi\alpha z^4}
\frac{\epsilon(0)-1}{\epsilon(0)+1} 
\equiv  -\frac{C_4}{z^4} \,.
\end{equation}
but this result cannot be used for the case of conductors, where $\epsilon$ has a pole
at $\omega$=0. We have checked that when the $\epsilon(\ii\omega)$ (a real function)
is bounded from below by $a/\omega$, as it is in the case of conductors, the long
distance behavior of the potential is equal to the one of perfect conductor,
asymptotically parametrized by $C_4^{(\infty)}$:
\begin{equation}
V^{\rm cond}_1(z) \mathop{\sim}^{z\to\infty}V^{(\infty)}_1(z) \mathop{\sim}^{z\to\infty}-
\frac{3\,\alpha_1(0)}{8\pi\alpha z^4}\equiv-\frac{C_4^{(\infty)}}{z^4} \,.
\end{equation}

\section{Frequency dependent dielectric permittivity of gold}
The {\em ab-initio} computation of the frequency dependent dielectric 
response function of metals is beyond the reach of present day electronic
structure calculations.
In this work, the complex frequency dependent permittivity
of gold was reconstructed from the experimental optical data.
The imaginary part of the dielectric constant was modeled using a function:
\begin{equation}\label{model1}
{\rm Im}\,\epsilon(\omega)=
\frac{\omega_{\rm p}^2\,\omega_\tau}{\omega (\omega^2+\omega^2_\tau)}
+\sum_n c_n\,{\rm f}_{\rm TL}(\omega_{0,n},\omega'_{0,n},\gamma_n;\omega)\,,
\end{equation}
where the first term is the Drude model of the free-electron contribution, and
the second is a sum of empirical Tauc-Lorentz\cite{JeMo1996} functions:
\begin{equation}
\label{tauclorenz}
{\rm f}_{\rm TL}(\omega_0,\omega'_0,\gamma;\omega)=
\frac{\omega_0\,\gamma\,(\omega-\omega_0')^2\,\theta(\omega-\omega_0)}{\omega\,(\omega^2 - \omega_0^2)^2 + \gamma^2\,\omega^2}\,,
\end{equation}
where $\theta(x)$ is the Heaviside step function [$\theta(x)=0$ for $x<0$, and
$\theta(x)=0$ otherwise].  Once a satisfactory representation of the imaginary
part of $\epsilon(\omega)$ is found, the real part of the permittivity can be
calculated using the Kramers-Kroenig relation.  The parameters of
Eq.~(\ref{model1}) have been fitted to the experimental optical data. The
data used included the data sets collected in the handbook by
Palik\cite{Pa1985} which cover the visible, ultraviolet and X-ray regimes, and
various compilations of optical data in the microwave and in the infrared
regions of the electromagnetic spectrum\cite{AuData}.  

\begin{figure} 
\begin{center}
\psfig{file=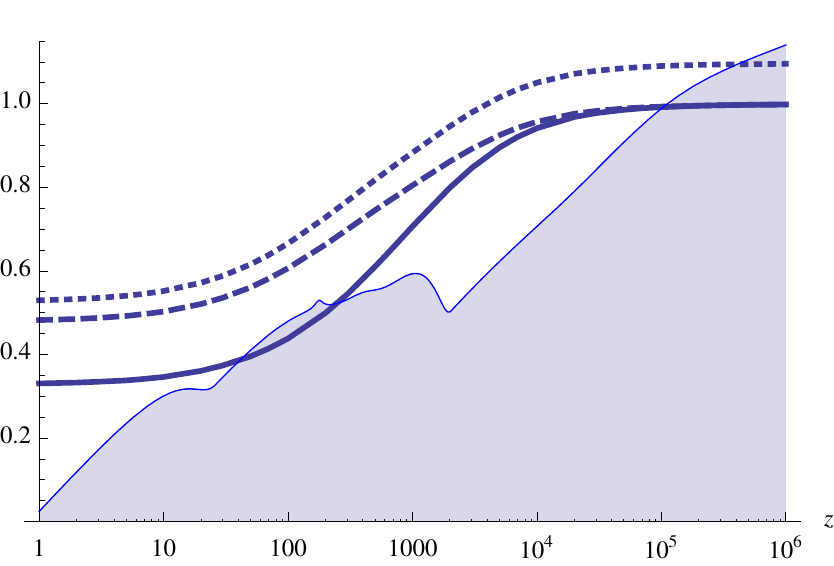,width=3in} 
\caption{Ratio between the full He-Au potential $V_1(z)$ calculated using the
optical data in~(\ref{model1}) and $V_1^{(\infty)}(z)$ (solid curve), between $V_1(z)$
and the approximate model~(\ref{simple}) using $C_4=C_4^{(\infty)}$ (dashed curve)
or using the best fit value of $C_4=44$~eV\AA$^4$ (dotted curve).
The shaded spectrum in the background depicts $\log{\rm Im}\,\epsilon(\omega)$
of the fit obtained in Table~\ref{table1}, for $\omega=c/z$. \label{fig1}}
\label{aba:fig1} 
\end{center}
\end{figure}

The results of a global fit of the model~(\ref{model1}) to the experimental data for
gold are listed (in atomic units) in Table~\ref{table1} and depicted in Fig~\ref{fig1}.
Our fitted values of the plasma frequency ($\omega_{\rm p}$), and the damping
frequency ($\omega_\tau$) can be compared to the values used by Lambrecht
{\em et al.}\cite{La2006} who used values of  $\omega_{\rm p}$=0.330, $\omega_\tau$=0.00108
and $\omega_{\rm p}$=0.276, $\omega_\tau$=0.00478.

\begin{table}
\tbl{Best fit parameters for the analytic model of ${\rm Im}\,\epsilon(\omega)$ for gold
according to Eq.~(\ref{model1}).\label{table1}}
{\begin{tabular}{@{} c D{,}{.}{6} c D{,}{.}{6} c D{,}{.}{6} c D{,}{.}{6} @{}}\toprule
 $\omega_{\rm p}$ & \multicolumn{3}{c}{0.357091} & $\omega_\tau$  & \multicolumn{3}{c}{0.001636}  \\
\hline
 $c_1$ & 3,177115 & $\omega_1$ & 0,117730 & $\omega'_1$ & 0,061100  &  $\gamma_1$  &  0,114560  \\
 $c_2$ & 0,483874 & $\omega_2$ & 0,337153 & $\omega'_2$ & \multicolumn{1}{c}{$=\omega'_1$} & $\gamma_2$ &  0,262558  \\
 $c_3$ & 0,106614 & $\omega_3$ & 0,777090 & $\omega'_3$ & \multicolumn{1}{c}{$=\omega'_1$} & $\gamma_3$ &  0,140060  \\
 $c_4$ & 1,988346 & $\omega_4$ & 1,014223 & $\omega'_4$ & \multicolumn{1}{c}{$=\omega'_1$} & $\gamma_4$ &  2,017446  \\
 $c_5$ & 2,220095 & $\omega_5$ & 5,242422 & $\omega'_5$ & \multicolumn{1}{c}{$=\omega_5$} & $\gamma_5$ & 10,076456 \\
\botrule
\end{tabular}}
\end{table}

When using the model defined in Eq.~(\ref{model1}), we observe a slight, but systematic deviation 
of the fitted function with respect to the  experimental data for the imaginary part of
the permittivity in the very low frequency region. In order to improve the fit, we decided to
further refine our model for the dielectric function by including an additional Drude-model
term, corresponding to a conductor with two types of carriers:
\begin{equation}
\label{model2}
{\rm Im}\,\epsilon(\omega)=\frac{\omega_{p_1}^2\,\omega_{\tau_1}}{\omega (\omega^2+\omega^2_{\tau_1})}
+\frac{\omega_{p_2}^2\,\omega_{\tau_2}}{\omega (\omega^2+\omega^2_{\tau_2})}
+\sum_n c_n\,{\rm f}_{\rm TL}(\omega_n,\omega'_n,\gamma_n;\omega)\ .
\end{equation}
The parameter of the model function have been reoptimized using nonlinear least squares fit
to the optical data, and their values are presented in Table~\ref{table2}.
\begin{table}
\tbl{Best fit parameters for the two plasma frequency model of ${\rm Im}\,\epsilon(\omega)$ for gold
according to Eq.~(\ref{model2}).\label{table2}}
{\begin{tabular}{@{} c D{,}{.}{6} c D{,}{.}{6} c D{,}{.}{6} c D{,}{.}{6} @{}}\toprule
 $\omega_{p_1}$ & 0,327756 & $\omega_{\tau_1}$ & 0,001127 & $\omega_{p_2}$ & 0,107482 & $\omega_{\tau_2}$ & 0,019638 \\
\hline
 $c_1$ & 4,084274 & $\omega_1$ & 0,110273 & $\omega'_1$ &  0,066774  &  $\gamma_1$  &  0,108472  \\
 $c_2$ & 0,478826 & $\omega_2$ & 0,337918 & $\omega'_2$ & \multicolumn{1}{c}{$=\omega'_1$} & $\gamma_2$ &  0,259896  \\
 $c_3$ & 0,108575 & $\omega_3$ & 0,776982 & $\omega'_3$ & \multicolumn{1}{c}{$=\omega'_1$} & $\gamma_3$ &  0,140290  \\
 $c_4$ & 2,001595 & $\omega_4$ & 1,011645 & $\omega'_4$ & \multicolumn{1}{c}{$=\omega'_1$} & $\gamma_4$ &  2,010662  \\
 $c_5$ & 2,193675 & $\omega_5$ & 5,168412 & $\omega'_5$ & \multicolumn{1}{c}{$=\omega_5$}  & $\gamma_5$ & 10,819556  \\
\botrule
\end{tabular}}
\end{table}
The double Drude model leads to a significant improvement of the low-frequency behavior of $\epsilon$, but has a negligible influence on the calculated potential. Since there is no physical justification for Eq.~(\ref{model2}), we use Eq.~(\ref{model1})
for the results obtained in the rest of this paper.

\section{Numerical Results}
The atom-surface interaction potentials have been computed by numerically evaluating
Eq.~(\ref{dipanydielectric}) using Eqs.~(\ref{model1}) and~(\ref{model2}) and {\em ab-initio}
calculated values of dynamic dipole polarizabilities. For the case of helium the respected
polarizabilities have been computed variationally using basis sets of explicitly correlated
functions\cite{La2004}. For the other atoms considered we used dynamic
polarizabilities published by  Derevianko, Porsev and Babb\cite{Ba2009}.

We find it convenient to represent the potential by its short distance limit
multiplied by a ``damping function'', accounting for the effects of retardation,
which for the dipole case reads:
\begin{equation}
\label{fz}
V_1(z)=-\frac{C_3}{z^3}f_3(z)\ ,
\end{equation}
For the damping function $f_3(z)$, we found the following functional form to lead
to an accurate representation of all interaction potential for all atom-wall distances:
\begin{equation} \label{dampingfun}
f_3(z)=\frac{1+a_1\,\alpha z+a_2 (\alpha z)^2}{1+a_1\,\alpha z+b_2 (\alpha z)^2+b_3 (\alpha z)^3}\ ,
\end{equation}
with $\alpha$ being the fine structure constant.  The equality of the linear terms
in both the numerator and in the denominator of this rational function is due to the
requirement of the potential having the correct short distance limit:
$f_3(z)=1+\mathcal{O}[(\alpha z)^2]$. The best values of the parameters $a_1$, $a_2$,
$b_2$, $b_3$ in~(\ref{dampingfun}) fitted to the calculated interaction potentials for
different atomic species are presented in Tables \ref{table3}-\ref{table5}, together
with the corresponding values of $C_3$.

\begin{table}
\tbl{Long range potentials for noble gas atoms interacting with a surface of gold.
The second row contains the $C_3$ constants, and the following rows
give the values of the $f_3(z)$ damping function, defined in Eq.~(\ref{fz})
for different distances. In the last four rows, we present the best fit parameters for the rational
function~(\ref{dampingfun}) which covers all distances.\label{table3}}
{\begin{tabular}{@{} c D{,}{.}{6} D{,}{.}{6} D{,}{.}{6} D{,}{.}{6} D{,}{.}{6}
 @{}}\toprule
  & \multicolumn{1}{c}{He} & \multicolumn{1}{c}{Ne} & \multicolumn{1}{c}{Ar} & \multicolumn{1}{c}{Kr} & \multicolumn{1}{c}{Xe} \\
\hline
 \multicolumn{1}{c}{$C_3$}     & 0,062(3) & 0,127(1) & 0,415(4) & 0,588(5) & 0,870(6) \\
\Hline
$z$ &  \multicolumn{5}{c}{$f_3(z) = V(z)/(C_3/z^3)$} \\
\hline
1$\times10^1$ & 0,99108 & 0,98804 & 0,99323 & 0,99348 & 0,99405 \\
2$\times10^1$ & 0,97388 & 0,96653 & 0,98014 & 0,98112 & 0,98275 \\
5$\times10^1$ & 0,90949 & 0,89155 & 0,92931 & 0,93352 & 0,93907 \\
1$\times10^2$ & 0,80490 & 0,77796 & 0,84192 & 0,85135 & 0,86320 \\
2$\times10^2$ & 0,64602 & 0,61531 & 0,69933 & 0,71522 & 0,73546 \\
5$\times10^2$ & 0,40337 & 0,37942 & 0,45908 & 0,47873 & 0,50522 \\
1$\times10^3$ & 0,25046 & 0,23481 & 0,29294 & 0,30931 & 0,33232 \\
2$\times10^3$ & 0,14492 & 0,13575 & 0,17185 & 0,18274 & 0,19849 \\
5$\times10^3$ & 0,06539 & 0,06124 & 0,07800 & 0,08322 & 0,09088 \\
1$\times10^4$ & 0,03441 & 0,03222 & 0,04109 & 0,04387 & 0,04796 \\
2$\times10^4$ & 0,01770 & 0,01657 & 0,02114 & 0,02257 & 0,02469 \\
5$\times10^4$ & 0,00721 & 0,00675 & 0,00861 & 0,00919 & 0,01005 \\
1$\times10^5$ & 0,00363 & 0,00339 & 0,00433 & 0,00462 & 0,00506 \\
\Hline
$a_1$     & 4,19366 & 5,91822 & 3,76881 & 3,95140 & 4,21331 \\
$a_2$     & 0,29208 & 0,47236 & 0,21430 & 0,20804 & 0,20365 \\
$b_2$     & 2,13922 & 3,33497 & 1,53023 & 1,47551 & 1,41033 \\
$b_3$     & 0,10861 & 0,18758 & 0,06671 & 0,06066 & 0,05430 \\
\botrule
\end{tabular}}
\end{table}

\begin{table}
\tbl{Long range potentials for alkali atoms interacting with a surface of gold.
The second row contains the $C_3$ constants, and the following rows
give the values of the $f_3(z)$ damping function, defined in Eq.~(\ref{fz})
for different distances. In the last four rows, we present the best fit parameters for the rational
function~(\ref{dampingfun}) which covers all distances.\label{table4}}
{\begin{tabular}{@{} c D{,}{.}{6} D{,}{.}{6} D{,}{.}{6} D{,}{.}{6} D{,}{.}{6}
 @{}}\toprule
  & \multicolumn{1}{c}{Li} & \multicolumn{1}{c}{Na} & \multicolumn{1}{c}{K} & \multicolumn{1}{c}{Rb} & \multicolumn{1}{c}{Cs} \\
\hline
 \multicolumn{1}{c}{$C_3$}     & 1,210(5) & 1,356(6) & 2,058(9) & 2,31(1) & 2,79(1) \\
\Hline
$z$ &  \multicolumn{5}{c}{$f_3(z) = V(z)/(C_3/z^3)$} \\
\hline
1$\times10^1$ & 0,99899 & 0,99845 & 0,99829 & 0,99801 & 0,99780 \\
2$\times10^1$ & 0,99740 & 0,99590 & 0,99542 & 0,99465 & 0,99408 \\
5$\times10^1$ & 0,98992 & 0,98533 & 0,98333 & 0,98086 & 0,97903 \\
1$\times10^2$ & 0,97415 & 0,96515 & 0,96065 & 0,95541 & 0,95169 \\
2$\times10^2$ & 0,94032 & 0,92509 & 0,91803 & 0,90850 & 0,90210 \\
5$\times10^2$ & 0,84866 & 0,82359 & 0,82012 & 0,80456 & 0,79521 \\
1$\times10^3$ & 0,73176 & 0,69946 & 0,70857 & 0,69126 & 0,68309 \\
2$\times10^3$ & 0,57369 & 0,53721 & 0,56295 & 0,54776 & 0,54485 \\
5$\times10^3$ & 0,34082 & 0,30970 & 0,34388 & 0,33492 & 0,33918 \\
1$\times10^4$ & 0,19776 & 0,17688 & 0,20301 & 0,19808 & 0,20323 \\
2$\times10^4$ & 0,10552 & 0,09370 & 0,10926 & 0,10672 & 0,11029 \\
5$\times10^4$ & 0,04353 & 0,03854 & 0,04525 & 0,04422 & 0,04585 \\
1$\times10^5$ & 0,02196 & 0,01943 & 0,02285 & 0,02233 & 0,02318 \\
\Hline
$a_1$      & 6,96178 & 19,81680 & 1,50957 & 1,90225 & 2,21418 \\
$a_2$      & 0,10676 &  0,30362 & 2,97205 & 2,06782 & 1,55523 \\
$b_2$      & 0,44672 &  1,43152 & 3,15260 & 2,30121 & 1,83158 \\
$b_3$      & 0,00659 &  0,02118 & 0,15449 & 0,11044 & 0,07999 \\
\botrule
\end{tabular}}
\end{table}

\begin{table}
\tbl{Long range potentials for alkaline atoms interacting with a surface of gold.
The second row contains the $C_3$ constants, and the following rows
give the values of the $f_3(z)$ damping function, defined in Eq.~(\ref{fz})
for different distances. In the last four rows, we present the best fit parameters for the rational
function~(\ref{dampingfun}) which covers all distances.\label{table5}}
{\begin{tabular}{@{} c D{,}{.}{6} D{,}{.}{6} D{,}{.}{6} D{,}{.}{6} D{,}{.}{6}
 @{}}\toprule
  & \multicolumn{1}{c}{Be} & \multicolumn{1}{c}{Mg} & \multicolumn{1}{c}{Ca} & \multicolumn{1}{c}{Sr} & \multicolumn{1}{c}{Ba} \\
\hline
 \multicolumn{1}{c}{$C_3$}     & 0,650(5) & 1,067(6) & 1,82(1) & 2,19(1) & 2,75(1) \\
\Hline
$z$ &  \multicolumn{5}{c}{$f_3(z) = V(z)/(C_3/z^3)$} \\
\hline
1$\times10^1$ & 0,99797 & 0,99790 & 0,99805 & 0,99788 & 0,99778 \\
2$\times10^1$ & 0,99388 & 0,99394 & 0,99442 & 0,99404 & 0,99382 \\
5$\times10^1$ & 0,97571 & 0,97722 & 0,97909 & 0,97810 & 0,97764 \\
1$\times10^2$ & 0,93887 & 0,94441 & 0,94957 & 0,94792 & 0,94755 \\
2$\times10^2$ & 0,86495 & 0,87903 & 0,89248 & 0,89054 & 0,89129 \\
5$\times10^2$ & 0,69195 & 0,72244 & 0,75855 & 0,75885 & 0,76468 \\
1$\times10^3$ & 0,51642 & 0,55531 & 0,61130 & 0,61588 & 0,62880 \\
2$\times10^3$ & 0,34021 & 0,37699 & 0,44043 & 0,44893 & 0,46827 \\
5$\times10^3$ & 0,16580 & 0,18850 & 0,23484 & 0,24315 & 0,26173 \\
1$\times10^4$ & 0,08878 & 0,10168 & 0,12963 & 0,13511 & 0,14765 \\
2$\times10^4$ & 0,04590 & 0,05270 & 0,06773 & 0,07078 & 0,07784 \\
5$\times10^4$ & 0,01872 & 0,02151 & 0,02772 & 0,02899 & 0,03196 \\
1$\times10^5$ & 0,00942 & 0,01082 & 0,01395 & 0,01460 & 0,01610 \\
\Hline
$a_1$     & 3,00255 & 3,88303 & 12,61180 & 3,09874 & 2,83515 \\
$a_2$     & 0,12664 & 0,12802 &  0,19042 & 4,88813 & 2,70340 \\
$b_2$     & 0,51498 & 0,55406 &  1,27280 & 5,20283 & 3,03090 \\
$b_3$     & 0,01818 & 0,01601 &  0,01848 & 0,41434 & 0,20650 \\
\botrule
\end{tabular}}
\end{table}

The potentials presented above can be compared to simplified ones used
previously\cite{DrDK2003} in the analysis of the atomic beam experiment
measuring the quantum reflectivity. There, retardation has been accounted
for using:
\begin{equation}\label{simple}
V(z)=-\frac{C_4}{z^3(z+\lambdabar)}\ ,
\end{equation}
in which $\lambdabar$ represents the reduced wavelength of the first electronic
dipole transition in He, i.e. $\lambdabar=178\,$Bohr (93\AA). This
function has also been used in the preliminary evaluation of the experimental
results for $^3$He atoms quantum reflecting from a single crystal gold surface.
Here however, significant deviations have been found indicating, that
the coarse-grained model~(\ref{simple}) is not sufficiently complex to mimic
the more involved rational structure given in Eqs.~(\ref{fz}) and~(\ref{dampingfun}).

An independent check of our potentials has been performed by using our model
for the dielectic constant $\epsilon(\ii\omega)$ of gold to calculate the interaction
energy between two surfaces of gold, for both semi-infinite solids and for thin foils.
Our computations are within $1\%$ of the revisited results of Lambrecht {\em et al}\cite{La2007,La2008}.
This gives us confidence, that the ready-to-use interaction potentials for the
various atomic species presented in this paper are accurate. In addition, it
demonstrates that our simple analytic model of the permittivity of gold may
be very useful for future computation of Casimir potential involving gold, for both
microscopic and macroscopic bodies.

%
%
\section*{Acknowledgements}

This project was supported by the
National Science Foundation
(Grant PHY--8555454) and by a precision measurement grant from the
National Institute of Standards and Technology.
G.L.~acknowledges support by the Deutsche
Forschungsgemeinschaft (DFG, contract Je285/5--1).

\bibliographystyle{ws-procs9x6}
\bibliography{ws-pro-sample}

\end{document}